\documentstyle[referee]{laa}

\begin{document}
\thesaurus{07  
	   (07.09.1; 
	   07.13.1;  
	   07.13.2   
	   )}

\title{ON THE POYNTING-ROBERTSON EFFECT AND ANALYTICAL SOLUTIONS}

\author{ Jozef Kla\v{c}ka }
\institute{Institute of Astronomy,
Faculty for Mathematics and Physics,
Comenius University, \\
Mlynsk\'{a} dolina,
842 28 Bratislava,
Slovak Republic}
\date{}
\maketitle
\begin{abstract}
Solutions of the two-body problem with the simultaneous action
of the solar electromagnetic radiation in the form of the Poynting-Robertson
effect are discussed. Special attention is devoted to pseudo-circular orbits
and terminal values of osculating elements. The obtained results complete
those of Kla\v{c}ka and Kaufmannov\'{a} (1992) and Breiter and Jackson (1998).

Terminal values of osculating elements presented in Breiter and Jackson (1998)
are of no physical sense due to the
fact that relativistic equation of motion containing only first order of
$\vec{v}/c$ was used in the paper.


\end{abstract}

\section{Introduction}
Breiter and Jackson (1998; BJ-paper in the following text) have presented analytical
mathematical solutions of two-body problems with drag. The Poynting-Robertson
effect (P-R effect) is one of the special cases discussed in the BJ-paper.

The conclusion of the BJ-paper concerns the fact that analytical solution
in terms of special functions exists for the P-R effect. However, this
result is useful only in the two limiting cases: pseudo-circular orbits
and terminal values of osculating elements. As for the general case, there is
a complication with calculation of infinite functional series and even
with the divergence of the series.

The problem of the divergence of the infinite functional series can be overcome
by numerical integration of the equation(s) of motion(s). This can be easily
done at the present epoch of computers. However, it is nice, indeed, when
a man can compare his numerical calculations with analytical solutions in the
limiting cases -- analytical solutions are very interesting.

The aim of this paper is to discuss analytical solutions for the P-R effect
in the two limiting cases: pseudo-circular orbits and terminal values of
osculating elements. It is shown that the real results do not completely
correspond
to the results presented in the BJ-paper. General analytical solutions
for pseudo-circular orbits
presented in our paper are presented.

\section{Overview of the correctly presented results}
The first correct special result for the P-R effect was presented by Robertson
(1937; cited in accordance with BJ-paper). The complete formula for the P-R effect
is correctly derived in Kla\v{c}ka (1992a). Other, more simple correct
derivations may be found in Kla\v{c}ka's papers: 1992b, 1993a, 1993b.
These papers present also arguments as for physical incorrectness in papers
cited in the BJ-paper (except of Robertson).

As for the solutions of the equation of motion for the P-R effect, we refer
also to Kla\v{c}ka (1992c; error is in Eq. (10) of the paper -- the
right-hand side of Eq. (10) must contain $\mu ~ ( 1 ~-~ \beta )$ instead of
$\mu$ when used in the section 3),
Kla\v{c}ka (1993c, 1993d), Kla\v{c}ka and Kaufmannov\'{a}
(1992, 1993 -- typewriting error in Eq. (1)) (see also Kla\v{c}ka 1994a).
As for other papers of the author, dealing mainly with
general interaction of the electromagnetic radiation (of the Sun; and, also,
solar corpuscular radiation), we refer to:
Kla\v{c}ka (1993f, 1993g (some numerical errors which may be easily found
are in the last section; moreover, real particle should rotate around one
axis -- axis of rotation), 1994b),
Kla\v{c}ka and Kocifaj (1994).

\section{P-R effect and analytical solutions}
At first, we put Eq. (18) in BJ-paper into a correct form:
\begin{equation}\label{1}
\dot{r} = \frac{\mu}{\alpha_{t}} ~ x^{\nu} ~ ( \nu ~ y ~+~ x ~ y' ) ~.
\end{equation}

In the case of the P-R effect we have $\nu =$ 1, $\mu$ must be substituted
by $\mu_{0} ~( 1 ~-~ \beta )$ and
$\alpha_{t} = \beta ~ \mu_{0} ~/~ c$:
\begin{eqnarray}\label{2}
\dot{r} &=& c~ \frac{1~-~\beta}{\beta} ~ x ~ ( y ~+~ x ~ y' ) ~,
 \nonumber \\
\alpha_{t} &=& \beta ~ \mu_{0} ~/~ c ~.
\end{eqnarray}

\subsection{Initial conditions}
Let the initial orbit is given by $r_{in}$, $\dot{r}_{in}$, $h_{in}$ and
$\vartheta_{in} =$ 0. Eqs. (5) in BJ-paper and Eqs. (2) of our paper yield
\begin{eqnarray}\label{3}
x_{in} &=& \frac{c}{\beta ~ \mu_{0}} ~ h_{in} ~, \nonumber \\
y_{in} &=& \frac{h_{in}^{2}}{\mu_{0} ~ ( 1 ~-~ \beta ) ~ r_{in}} ~ x_{in}^{-~3} ~,
\nonumber \\
y_{in} ' &=& \left \{ \frac{\dot{r}_{in}}{c} ~ \frac{\beta}{1 ~-~ \beta} ~ x_{in}^{2}
~-~ \frac{h_{in}^{2}}{\mu_{0} ~ ( 1 ~-~ \beta ) ~ r_{in}} \right \}
 ~ x_{in}^{-~4} ~.
\end{eqnarray}
These equations complete Eqs. (16) in BJ-paper.

\section{Special types of pseudo-circular orbits}

\subsection{Increasing eccentricity}
Due to the fact that Bessel's functions $J_{1} (x)$, $Y_{1} (x)$ can be expressed
as linear combinations of $\cos(x ~-~ 3 ~ \pi ~/~ 4)$,
$\sin(x ~-~ 3 ~ \pi ~/~ 4)$, Eqs. (16) in BJ-paper enable to fulfil the conditions
$A = B =$ 0. These conditions are fulfiled by conditions
\begin{eqnarray}\label{4}
y_{in} &=& \hat{S}_{1} ( x_{in} ) ~,
\nonumber \\
y_{in} ' &=& \hat{S}_{1} ' ( x_{in} ) ~.
\end{eqnarray}

Eqs. (3) and (4) lead to special type of pseudo-circular orbits,
as it is discussed in section
5 in BJ-paper. The procedure goes in the way that giving $h_{in}$ we can
calculate $x_{in} = h_{in} ~c~ / ( \beta ~ \mu_{0} )$; the first of Eqs. (4)
and second of Eqs. (3) yield
\begin{equation}\label{5}
r_{in} = \frac{\beta ^{3}}{1 ~-~ \beta} ~ \frac{\mu_{0}^{2}}{h_{in}} ~
\left \{  \hat{S}_{1} ( x_{in} ) \right \} ^{-~1} ~;
\end{equation}
the second of Eqs. (4) and the third of Eqs. (3) yield
\begin{equation}\label{6}
\dot{r}_{in} = c ~ \frac{1 ~-~ \beta}{\beta} ~ x_{in}^{-2} ~
\left \{ x_{in}^{4} ~\hat{S}_{1} ' ( x_{in} ) ~+~
\frac{h_{in}^{2}}{\mu_{0} ~ ( 1 ~-~ \beta ) ~ r_{in}} \right \} ~.
\end{equation}
Transversal component of velocity may be calculated from
the relation $v_{T ~ in} = h_{in} ~/~ r_{in}$.

The case $e \approx 2 / x$ (Eq. (28) in BJ-paper), together with
$\sqrt{\mu_{0} ~ ( 1 ~-~ \beta ) ~ p}$ $=$ $x~ \beta ~ \mu_{0} ~/~ c$, leads to
\begin{equation}\label{7}
p~ e^{2} \approx  \frac{\left ( 2~ \beta \right )^{2}}{1 ~-~ \beta} ~
\frac{\mu_{0}}{c^{2}} ~.
\end{equation}

In any case the conclusion about the increasing eccentricity is not
completely correct. The eccentricity begin to oscilate at some state of the
orbital evolution, although its mean value is an increasing function
of time. The conclusion ``The general conclusion is that in the
pseudo-circular solution the osculating eccentricity {\it grows systematically}
from the small but nonzero value of order $\alpha_{t}/h$ towards $e =$ 1.''
(BJ-paper, section 5) is incorrect (see also section 4 of our paper).

\subsection{Oscillating eccentricity}
Pseudo-circular orbit discussed in BJ-paper is only a very special case
of possible pseudo-circular orbits. We complete the case by all the other
possibilities, where also zero value of eccentricity is possible.

Let us consider a situation when meteoroid is ejected from comet at cometary
aphelion with zero ejection velocity. If the comet's orbital elements are
$a_{0}$, $e_{0}$ and meteoroid's $\beta$ is given by the condition
$\beta = e_{0}$, then Eqs. (30) and (31) in Kla\v{c}ka (1992c) and
Eq. (17) in Kla\v{c}ka (1993e;
Eq. (17) must contain $\mu_{0} ~ ( 1 ~-~ \beta )$ instead of
$\mu$, now) yield
\begin{eqnarray}\label{8}
e_{in} &=& 0 ~, \nonumber \\
a_{in} &=& a_{0} ~ ( 1 ~+~ e_{0} ) ~, \nonumber \\
H_{in} &=& \sqrt{\mu_{0} ~ a_{0} ~ ( 1 ~-~ e_{0}^{2} )} ~.
\end{eqnarray}
(If meteoroid is ejected with nonzero velocity, one can use equations
of Gajdo\v{s}\'{\i}k and Kla\v{c}ka (1999).) Then, we have
\begin{eqnarray}\label{9}
r_{in} &\equiv& r_{0} = a_{0} ~ ( 1 ~+~ e_{0} ) ~, \nonumber \\
\dot{r}_{in} &\equiv& \dot{r}_{0} = 0 ~.
\end{eqnarray}
If we set $\vartheta_{0} = 0$, $\alpha_{t} = \beta ~ \mu_{0} ~/~ c$, equations
of sections 2, 3 and 4 of BJ-paper  yield
\begin{eqnarray}\label{10}
h &=& \sqrt{\mu_{0} ~ a_{0} ~ ( 1 ~-~ e_{0}^{2} )} ~, \nonumber \\
x_{in} &\equiv& x_{0} = h ~ \alpha_{t}^{-1} = ( c~/~e_{0} ) ~
	  \sqrt{\mu_{0} ~ ( 1 ~-~ e_{0}^{2} ) ~/~a_{0}}  ~, \nonumber \\
y_{in} &\equiv& y_{0} = \{ \mu_{0} ~ ( 1 ~-~ \beta ) ~ r_{in} ~ x_{0} \}^{-1} ~ \alpha_{t}^{2} =
	  x_{0}^{-3} ~.
\end{eqnarray}
Eq. (2) yields for $\dot{r}_{in} = 0$:
$y' = - ~ y ~/~ x$,  and, thus
\begin{equation}\label{11}
y_{in} ' \equiv y_{0} ' = - ~ y_{0} ~/~ x_{0} = -~ x_{0}^{-4} ~.
\end{equation}
This case should correspond to the case of
Kla\v{c}ka and Kaufmannov\'{a} (1992, 1993). We do not treat it here, since
general pseudo-circular orbit will be discussed in the following section.

\section{General pseudo-circular orbit}
In order of generalizing of Eqs. (10) and (11),
we use Eqs. (5) of BJ-paper:
\begin{eqnarray}\label{12}
x_{in} &\equiv& x_{0} = \frac{c}{\beta ~ \mu_{0}} ~
	 \sqrt{\mu_{0} ~ ( 1 ~-~ \beta ) ~ a_{in} ~ ( 1 ~-~ e_{in}^{2} )} ~,
	 \nonumber \\
y_{in} &\equiv& y_{0} = \frac{\beta ^{2} ~ \mu_{0}}{c^{2} ~ ( 1 ~-~ \beta ) ~x_{0}} ~
	 \frac{1 ~+~ e_{in} ~ \cos f_{in}}{a_{in} ~ ( 1 ~-~ e_{in}^{2} )}
       = \frac{1 ~+~ e_{in} ~ \cos f_{in}}{x_{0}^{3}} \approx
	  \frac{1}{x_{0}^{3}}
\end{eqnarray}
for pseudo-circular orbits.
Eq. (3) yields
\begin{equation}\label{13}
y'_{0} \approx \left \{ \frac{\sqrt{1 ~-~ \beta}}{\beta}
	       ~ \frac{c}{\sqrt{\mu_{0} / r_{in}}} ~ e_{in} ~ \sin f_{in}
	       ~-~ 1 \right \} ~ x_{in}^{-4} \equiv -~ \frac{k}{x_{0}^{4}}
\end{equation}
for pseudo-circular orbits;
$y'_{0}$ $=$ $-~ 1~/~ x_{0}^{4}$ for $\dot{r}_{in} =$ 0 (Eq. (11)),
$y'_{0}$ $=$ $-~ 3~/~ x_{0}^{4} ~+~ ...$ for $A = B =$ 0 (Eq. (4)).

The second of Eqs. (12) and Eq. (13) yield
\begin{eqnarray}\label{14}
A &\approx&  -~ \sqrt{\pi ~/~ 2} ~ x_{0}^{-7/2} ~
	    ( 3 ~-~ k )~ \sin ( x_{0} ~-~  3 ~\pi ~/~ 4 ) ~+~
\nonumber \\
  & &	     +~ \sqrt{\pi ~/~ 2} ~ x_{0}^{-7/2} ~
	   \frac{55~+~ 3 ~ k}{8 ~ x_{0}} ~ \cos ( x_{0} ~-~  3 ~\pi ~/~ 4 ) ~,
\nonumber \\
B &\approx& +~ \sqrt{\pi ~/~ 2} ~ x_{0}^{-7/2} ~
	    ( 3 ~-~ k )~ \cos ( x_{0} ~-~  3 ~\pi ~/~ 4 ) ~+~
\nonumber \\
  & &	     +~ \sqrt{\pi ~/~ 2} ~ x_{0}^{-7/2} ~
	   \frac{55~+~ 3 ~ k}{8 ~ x_{0}} ~ \sin ( x_{0} ~-~  3 ~\pi ~/~ 4 ) ~.
\end{eqnarray}

If $x = h ~ \alpha_{t}^{-1} ~-~ \vartheta$, then
\begin{eqnarray}\label{15}
Z_{1}(x) &=& A~ J_{1}(x) ~+~ B~ Y_{1}(x)
	 \approx
	   x_{0}^{-7/2} ~ x^{-1/2}  ~
	   ( 3 ~-~ k )~ \sin ( x ~-~ x_{0} ) ~+~ \nonumber \\
  & &	   x_{0}^{-7/2} ~ x^{-1/2}
	 \left \{ \frac{9 ~-~ 3~k}{8~ x} ~+~ \frac{55 ~+~ 3~k}{8~ x_{0}}
	 \right \} ~
	 \cos ( x ~-~ x_{0} )  ~, \nonumber \\
\hat{S}_{1}(x) &\approx& x^{-3} ~-~ 8~ x^{-5}  ~, \nonumber \\
Z_{0}(x) &=& A~ J_{0}(x) ~+~ B~ Y_{0}(x)
	 \approx
	   x_{0}^{-7/2} ~ x^{-1/2} ~
	   ( 3 ~-~ k )~ \cos ( x ~-~ x_{0} ) ~+~ \nonumber \\
  & &	   x_{0}^{-7/2} ~ x^{-1/2}
	 \left \{ \frac{3 ~-~ k}{8~ x} ~-~ \frac{55 ~+~ 3~k}{8~ x_{0}}
	 \right \} ~
	 \sin ( x ~-~ x_{0} )  ~, \nonumber \\
\hat{S}'_{1}(x) &\approx& ~-~ 3~x^{-4} ~+~ 40~ x^{-6}  ~.
\end{eqnarray}
One obtains, then
\begin{eqnarray}\label{16}
e ~ \cos f  &=&  x^{3} ~ ( Z_{1}(x) ~+~ \hat{S}_{1}(x) )   ~-~ 1 \approx
		  \nonumber \\
	    &\approx&
	   ( 3 ~-~ k ) ~x_{0}^{-7/2} ~ x^{5/2} ~ \sin ( x ~-~ x_{0} ) ~-~
	   8 ~ x^{-2} ~+~ \cos ( x ~-~ x_{0} ) \times \nonumber \\
    & &    \left \{
\frac{9 ~-~3~ k}{8} ~ ~x_{0}^{-7/2} ~ x^{3/2} ~+~
	   \frac{55 ~+~3~ k}{8} ~ ~x_{0}^{-9/2} ~ x^{5/2} \right \}
~, \nonumber \\
e ~ \sin f  &=&  x^{3} ~ ( Z_{0}(x) ~+~ x^{-1} ~ \hat{S}_{1}(x) ~+~
		 \hat{S}'_{1}(x) ) \approx
		  \nonumber \\
	    &\approx&
	   ( 3 ~-~ k ) ~x_{0}^{-7/2} ~ x^{5/2} ~ \cos ( x ~-~ x_{0} ) ~-~
	   2 ~ x^{-1} ~+~ \sin ( x ~-~ x_{0} ) \times \nonumber \\
    & &    \left \{
\frac{3 ~-~ k}{8} ~ ~x_{0}^{-7/2} ~ x^{3/2}  ~-~
	   \frac{55 ~+~3~ k}{8} ~ ~x_{0}^{-9/2} ~ x^{5/2} \right \}
~, \nonumber \\
e^{2} &\approx& 4 ~ x^{-2} ~+~ ( 3 ~-~ k )^{2} ~ x_{0}^{-7} ~ x^{5}  ~-~
	  4~ ( 3 ~-~ k ) ~x_{0}^{-7/2} ~ x^{3/2} ~ \cos ( x ~-~ x_{0} )
   \nonumber \\
 & & +~ x_{0}^{-7/2} ~ x^{1/2} ~ \sin ( x ~-~ x_{0} ) ~ \times	 \nonumber \\
 & &
  \left \{ -~ \frac{33~ ( 3 ~-~ k )}{2} ~-~ \frac{55~+~3~k}{2} ~x_{0}^{-1} ~ x
 ~+~ ( 3 ~-~ k ) ~ \cos ( x ~-~ x_{0} ) \times \right .
   \nonumber \\
 & & \left .
   \left [ \frac{15~-~5~k}{4} \left ( x_{0}^{-1} ~ x \right ) ^{7/2}
  ~-~ \frac{55~+~3~k}{4} \left ( x_{0}^{-1} ~ x \right ) ^{9/2} \right ]
  \right \} ~.
\end{eqnarray}
The equation for $e^{2}$ represents general type of evolution of eccentricity
for pseudo-circular orbit in comparison with
Eq. (28) in BJ-paper and papers by
Kla\v{c}ka and Kaufmannov\'{a} (1992, 1993).

\subsection{Time evolution of eccentricity and true anomaly}
Eqs. (16) enable us to investigate time evolution of osculating
eccentricity and true anomaly. We must bear in mind that the term
`increasing time' corresponds to the term `decreasing $x$'.

\subsubsection{Extremes of eccentricity}
The local extremes of eccentricity as a function of time are given by
condition (very large values of $x_{0}$ are considered, theoretically
$x_{0} \rightarrow \infty$; $k \ne$ 3)
\begin{equation}\label{17}
\frac{d e^{2}}{d x} =  0  ~ \Longleftrightarrow ~ \sin ( x ~-~ x_{0} )
			      \approx 0 ~.
\end{equation}
As for the second derivative, we have
\begin{equation}\label{18}
\frac{d^{2} e^{2}}{d x^{2}} \approx  4 ~( 3 ~-~ k ) ~ x_{0}^{-7/2} ~x^{3/2} ~
				     \cos ( x ~-~ x_{0} ) ~.
\end{equation}

The case $k = 3$ yields no local extreme for the
leading terms. Eqs. (16) yield for evolution of true anomaly $f$:
3 $\pi$ / 2 $\rightarrow$ $\pi$ and the value $\pi$ never occurs.

The case $k =$ 1 cannot be treated analytically, as it will be shown later on.

\subsubsection{$x=x_{0}$}
Putting $x=x_{0}$ into Eqs. (16), (17) and (18), one obtains that the point
$x=x_{0}$ corresponds to:
i) local minimum for the case $k <$ 3,
ii) local maximum for the case $k >$ 3,
for the function $e(t)$, or, $e(x)$.

If we want to find time evolution of true anomaly, we have to use Eqs. (16).
The time $t = t_{0} ~+~ \Delta t$, for small positive $\Delta t$, corresponds
to $x = x_{0} ~-~ \Delta x$, where $\Delta x$ is a small positive quantity
(in radians). Eqs. (16) yield, then
\begin{equation}\label{19}
e ~ \cos f \approx -~ ( 3 ~-~ k ) ~\frac{\Delta x}{x_{0}} ~~,~~~~
e ~ \sin f \approx \frac{1~-~k}{x_{0}} ~.
\end{equation}
Eqs. (19) imply, on the basis of $e >$ 0 and $| e ~ \sin f | \gg$
$| e ~ \cos f |$:
\begin{eqnarray}\label{20}
\lim_{x \rightarrow x_{0}^{-}} f &=& \pi ~/~ 2 ~~~~~~,~~~  k < 1~,
							 \nonumber \\
\lim_{x \rightarrow x_{0}^{-}} f &=& 3 ~ \pi ~/~ 2 ~~~,~~~  k > 1 ~.
\end{eqnarray}
As for the value of eccentricity, the last of Eqs. (16) yields
\begin{equation}\label{21}
e^{2} ( x = x_{0} )  \approx ( 1 ~-~ k )^{2} ~ x_{0}^{-2} ~.
\end{equation}

\subsubsection{$x = x_{0} ~-~ \pi$}
Putting $x = x_{0} ~-~ \pi$ into Eqs. (16), (17) and (18), one obtains that
the point $x = x_{0} ~-~ \pi$
corresponds to:
i) local maximum for the case $k <$ 3,
ii) local minimum for the case $k >$ 3,
for the function $e(t)$, or, $e(x)$.
The value of the osculating eccentricity is
\begin{equation}\label{22}
e^{2} (x = x_{0} ~-~ \pi)  \approx   ( 5 ~-~ k )^{2} ~ x_{0}^{-2} ~+~
			   \pi ~ ( 5 ~k ~-~11 ) ~ ( 5 ~-~ k ) ~ x_{0}^{-3} ~.
\end{equation}

If we want to find time evolution of true anomaly, we have to use Eqs. (16).

The situation shortly before the extreme corresponds to
$x = x_{0} ~-~ \pi ~+~ \Delta x$, where $\Delta x$ is a small positive quantity
(in radians) shortly before the extreme, and,
$\Delta x$ is a small negative quantity
(in radians) shortly after the extreme. Eqs. (16) yield, then
\begin{equation}\label{23}
e ~ \cos f \approx -~ ( 3 ~-~ k ) ~\frac{\Delta x}{x_{0}} ~~,~~~~
e ~ \sin f \approx -~ \frac{5~-~k}{x_{0}} ~.
\end{equation}
Eqs. (23) imply
\begin{eqnarray}\label{24}
\lim_{x \rightarrow x_{0} ~-~ \pi} f &=&  3 ~ \pi ~/~ 2 ~~~~~~~~~~~~~
	~~~~~~~~~~ ,~~~  k < 5~,
							 \nonumber \\
\lim_{x \rightarrow x_{0} ~-~ \pi} f &=&  \pi ~/~ 2 ~~~~~~~~~~~~~~~~~
	~~~~~~~~~ ,~~~	k > 5~.
\end{eqnarray}

Eqs.(24) yield that true anomaly is a continuous
function at $x = x_{0} ~-~ \pi$ -- at the first local maximum of
osculating eccentricity.

\subsubsection{$x = x_{0} ~-~ 2~ \pi$}
Putting $x = x_{0} ~-~ 2~ \pi$ into Eqs. (16), (17) and (18), one obtains that the point
$x=x_{0} ~-~ 2~ \pi$ corresponds to:
i) local minimum for the case $k <$ 3,
ii) local maximum for the case $k >$ 3,
for the function $e(t)$, or, $e(x)$.
As for the value of eccentricity, the last of Eqs. (16) yields
\begin{equation}\label{25}
e^{2} ( x = x_{0} ~-~ 2 ~ \pi )  \approx ( 1 ~-~ k )^{2} ~ x_{0}^{-2}
		     ~+~ 2~ \pi ~ ( 5 ~k ~-~11 ) ~ ( 5 ~-~ k ) ~ x_{0}^{-3} ~.
\end{equation}
Comparison of Eqs. (21) and (25) leads to the conclusion
\begin{eqnarray}\label{26}
e ( x_{0} ) &=& e ( x_{0} ~-~ 2 ~ \pi ) ~, ~ k = 11 / 5, 5
						     \nonumber \\
e ( x_{0} ) &<& e ( x_{0} ~-~ 2 ~ \pi ) ~, ~ 11 / 5 < k < 5
						     \nonumber \\
e ( x_{0} ) &>& e ( x_{0} ~-~ 2 ~ \pi ) ~, ~
k < 1, 1 < k < 11 / 5, 5 < k
\end{eqnarray}
Eq. (25) shows that the case $k =$ 1 leads to inconsistencies:
$e ( x = x_{0} ~-~ 2 ~ \pi ) =$ 0 -- only the leading term can be considered.

If we want to find time evolution of true anomaly, we have to use Eqs. (16).

The situation near extreme corresponds to
$x = x_{0} ~-~ 2~ \pi ~+~ \Delta x$, where
$\Delta x$ is a small positive quantity
(in radians) shortly before the extreme, and,
$\Delta x$ is a small negative quantity
(in radians) shortly after the extreme. Eqs. (16) yield, then
\begin{equation}\label{27}
e ~ \cos f \approx  ( 3 ~-~ k ) ~\frac{\Delta x}{x_{0}} ~~,~~~~
e ~ \sin f \approx  \frac{1~-~k}{x_{0}} ~.
\end{equation}
Eqs. (27) imply, on the basis of $e >$ 0 and $| e ~ \sin f | \gg$
$| e ~ \cos f |$:
\begin{eqnarray}\label{28}
\lim_{x \rightarrow x_{0} ~-~ 2 ~ \pi} f &=& \pi ~/~ 2 ~~~~~~,~~~  k < 1~,
							 \nonumber \\
\lim_{x \rightarrow x_{0} ~-~ 2 ~ \pi} f &=& 3 ~ \pi ~/~ 2 ~~~,~~~  k > 1 ~.
\end{eqnarray}

\subsubsection{$x = x_{0} ~-~ 3~ \pi$}
Putting $x = x_{0} ~-~ 3~ \pi$ into the last of Eqs. (16) one obtains that
the point
$x=x_{0} ~-~3~ \pi$ corresponds to:
i) local maximum for the case $k <$ 3,
ii) local minimum for the case $k >$ 3,
for the function $e(t)$, or, $e(x)$.
The value of the osculating eccentricity is
\begin{equation}\label{29}
e^{2} (x = x_{0} ~-~3~ \pi)  \approx   ( 5 ~-~ k )^{2} ~ x_{0}^{-2} ~+~
			  3~ \pi ~ ( 5 ~k ~-~11 ) ~ ( 5 ~-~ k ) ~ x_{0}^{-3} ~.
\end{equation}
Comparison of Eqs. (22) and (29) leads to the conclusion
\begin{eqnarray}\label{30}
e ( x_{0} ~-~ \pi ) &=& e ( x_{0} ~-~ 3 ~ \pi ) ~, ~ k = 11 / 5, 5
						     \nonumber \\
e ( x_{0} ~-~ \pi ) &<& e ( x_{0} ~-~ 3 ~ \pi ) ~, ~ 11 / 5 < k < 5
						     \nonumber \\
e ( x_{0} ~-~ \pi ) &>& e ( x_{0} ~-~ 3 ~ \pi ) ~, ~
k \in ( 11 / 5, 5 ) ' ~.
\end{eqnarray}
As for time evolution of true anomaly, the results are analogous to
Eqs. (23) -- (24).

\subsubsection{Discussion}
The values of eccentricities are collected in Eqs. (26) and (30). The consequence
of these equations is that
mean values of eccentricity
during the corresponding periods exhibit similar properties. The cases
$k =$ 11~/5, $k =$ 5
yield the constant values of mean eccentricities during a long time evolution.

The case $k =$ 1 was treated in detail in
Kla\v{c}ka and Kaufmannov\'{a} (1992). The main results are presented
in Figs. 1, 3 and 8 in Kla\v{c}ka and Kaufmannov\'{a} (1992).

We can
collect the results for the cases $k =$ 1 and $k =$ 5 in
the statement that true anomaly is a discontinuous
function:
i) at $x = x_{0} ~-~ 2~ \pi ~m$, $m \in N$ for $k =$ 1,
i) at $x = x_{0} ~-~ \pi ~( 2~m ~-~1 )$, $m \in N$ for $k =$ 5.
We collect the results:
\begin{eqnarray}\label{31}
k &=& 1: ~ \lim_{x \rightarrow x_{0}^{-}} f = \pi ~, ~~~~~~~~~~~~~~~~~~
~\lim_{x \rightarrow ( x_{0} ~-~ \pi )} f =  3~ \pi ~/~ 2  ~,~
 \nonumber \\
k &=& 1: ~\lim_{x \rightarrow ( x_{0} ~-~ 2~ \pi )^{+}} f =  2 ~ \pi ~, ~~~~
~\lim_{x \rightarrow ( x_{0} ~-~ 2~ \pi )^{-}} f =  \pi ~,
\end{eqnarray}
\begin{eqnarray}\label{32}
k &=& 5: ~ \lim_{x \rightarrow x_{0}^{-}} f = 3~ \pi ~/~ 2 ~, ~~~~~~~~~~~~~~
~ \lim_{x \rightarrow ( x_{0} ~-~ \pi )^{+}} f = 2~ \pi  ~, \nonumber \\
k &=& 5: ~ \lim_{x \rightarrow ( x_{0} ~-~ \pi )^{-}} f =  \pi ~,~~~~~~~~~~
~ \lim_{x \rightarrow ( x_{0} ~-~ 2~ \pi )} f =  3~ \pi ~/~ 2 ~ .
\end{eqnarray}

\section{Terminal values of osculating elements}
As for the terminal values of osculating elements presented in BJ-paper,
the results may be collected in two important statements:
\begin{equation}\label{33}
\lim_{x \rightarrow 0^{+}}  f = \pi ~~~; ~~ \lim_{x \rightarrow 0^{+}}	e = 1 ~.
\end{equation}
Let us calculate other important quantities. The results are:
\begin{eqnarray}\label{34}
\lim_{x \rightarrow 0^{+}}  r &=& 0 ~~~; ~~ \lim_{x \rightarrow 0^{+}}	a = 0 ~,
\nonumber \\
\lim_{x \rightarrow 0^{+}}  v_{T} &=& 0 ~~~; ~~
\lim_{x \rightarrow 0^{+}}  v_{R} = -~ \frac{c}{2} ~ \frac{1 ~-~ \beta}{\beta} ~,
\nonumber \\
\lim_{x \rightarrow 0^{+}}  H &=& 0 ~~~; ~~
\lim_{x \rightarrow 0^{+}}  E = - ~ \infty ~,
\end{eqnarray}
where $r$ is particle's distance from the central point mass, $a$ -- semimajor
axis, $v_{T}$ -- transversal component of the velocity vector,
$v_{R} \equiv \dot{r}$ -- radial component of the velocity vector,
$H$ -- angular momentum, $E$ -- total energy of the particle with respect
to the central point mass.

The obtained results presented by Eqs. (33) and (34) yield important
inconsistencies. The osculating trajectory is parabola ($e =$ 1), the particle
is situated at apocenter ($f = \pi$) and the total energy is $E = -~ \infty$.
Normal result is that $E =$ 0 for the case $e =$ 1. So, there is something
wrong with the results, it seems.

The results given by Eqs. (33) and (34) are correct as for mathematical
point of view -- mathematical solutions of the discussed limits, based
on the mathematical solution presented in BJ-paper. However, we are not
interested in mathematics as the main theme. We are interested in physics.
Thus, the important inconsistencies presented by Eqs. (33) and (34) should
show that physics is not completely correct. Really, physics is incorrect.

It is a physical nonsense when a particle losses an unlimited energy within
a finite time.
The results presented by Eqs. (33) and (34) correspond to this nonphysical
situation. A particle spirals toward $r =$ 0 in a finite time and its
potential energy decreases in an unlimited value.

Eqs. (34) yield a hint how to put the discussed inconsistencies into
a correct physics.

Since
\begin{equation}\label{35}
\lim_{x \rightarrow 0^{+}}  v_{R} = -~ \frac{c}{2} ~ \frac{1 ~-~ \beta}{\beta}
< - ~c ~,
\end{equation}
for $0 < \beta <$ 1/3, we have to use complete form of the P-R effect --
relativistic effect (Kla\v{c}ka 1992a, Eq. (140)).

Eq. (35) yields that the form of the equation of motion
containing only first order of $v/c$
could be acceptable only for $0 \le 1 ~-~ \beta \ll$ 1 -- only in this case
the requirement $v \ll c$ holds. However, the third Kepler's law yields
$T^{2} =$ 4 $\pi^{2}$ $a^{3}$ $\{ \mu_{\beta = 0} ~ ( 1 ~-~ \beta ) \}^{-1}$
and
$\lim_{\beta \rightarrow 1^{-}}  T = \infty$. Thus, the situation
$0 \le 1 ~-~ \beta \ll$ 1 is not physically interesting. This situation
is evident also from Eq. (30) in Kla\v{c}ka (1992c):
$\lim_{\beta \rightarrow 1^{-}}  e_{in} >$ 1 -- no inspiralling toward the center
occurs.

The conclusion of this section states that the physics used in BJ-paper is not
competent to say something about the terminal values of osculating elements. Any
comparison of the statements for various initial conditions can be done only for
$r_{final} \gg r_{g} \equiv 2  G M / c^{2}$, $v \ll c$.
(We refer also to Kla\v{c}ka 1994c.)
As an example we may mention
the time of inspiralling toward the point mass center, based on
approximations in first order in $v/c$
-- the `,time of inspiralling'' corresponds to $r_{final} \ge$ 200 km
for the mass of the Sun. As for the averaged equations for osculating elements,
the condition (104) in Kla\v{c}ka 1992d must be fulfilled).


\section{Conclusions}
We have completed analytical formulae (expansions) for pseudo-circular
orbits for the Poynting-Robertson effect.

The statement
`The zero values of our arbitrary constants do not imply that the osculating
$e = 0$, but they match the case of the zero mean eccentricity ...'
(Breiter and Jackson 1998, section 5 on page 240) is incorect --
it is not possible that any continuous non-negative quantity
has zero mean value, unless it is identical zero.

Analytical results for more general pseudo-circular orbits than those discussed
in (Breiter and Jackson 1998), were obtained in this paper.

We have shown that terminal values of osculating elements lead to serious
inconsistencies caused by the fact that relativistic equations of motion
only in first order in $v/c$ were used in Breiter and Jackson (1998).

Finally, we have to stress several facts for elliptical orbits for
the P-R effect containing only first order in $v/c$
and in the zone of its applicability in two-body problem. \\
At first, the evident result is that osculating
semi-major axis is still a decreasing function of time	-- energy decreases
(see Eq. (22) in Kla\v{c}ka 1992c). \\
As for osculating eccentricity, it: \\
i) may be an increasing function of time for an initial long-time interval
if Eqs. (3) and (4) are fulfiled; \\
ii) still alternates in an increasing and a decreasing functions on
short-time intervals.\\
As for mean eccentricity, it may be: \\
i) an increasing function of time (for a suitable initial conditions:
see Fig. 3 in Kla\v{c}ka and Kaufmannov\'{a} (1992), or, Eqs. (3) and (4)
-- $k \in ( 11 ~/~ 5, 5 )$ for pseudocircular orbits); \\
ii) a constant function of time
(see Fig. 4 in Kla\v{c}ka and Kaufmannov\'{a} (1992); moreover,
the cases $k =$ 11~/~5, $k =$ 5 for pseudo-circular orbits); \\
iii) a decreasing function of time for some initial time interval followed
by an increasing function of time
(Kla\v{c}ka and Kaufmannov\'{a} (1992) --
$k \in ( 11~/~5, 5) '$ for pseudo-circular orbits); \\
iv) a decreasing function of time (Wyatt and Whipple 1950 -- without
derivation; correct derivation Kla\v{c}ka 1992c) for a long-time interval.   \\
(One must be careful which type of osculating elements is used.
The case ii) in mean eccentricity may corresponds to osculating elements
defined by value $\mu_{0}$ -- osculating elements I in Kla\v{c}ka 1992c,
or osculating elements in Kla\v{c}ka and Kaufmannov\'{a} (1992).
The cases i), iii), iv) and pseudo-circular orbits ii)
in mean eccentricity correspond to osculating
elements defined by value $\mu_{0} ~ ( 1 ~-~ \beta )$ -- osculating
elements II in Kla\v{c}ka 1992c, or, `non-osculating' elements in
Kla\v{c}ka and Kaufmannov\'{a} (1992).)

\acknowledgements
Special thanks to the firm ``Pr\'{\i}strojov\'{a} technika, spol. s r. o.''.
The paper was supported by the Scientific Grant Agency VEGA
(grants Nos. 1/4304/97 and 1/4303/97).

\end{document}